\documentclass[12pt]{iopart}
\usepackage{graphicx}
\begin{document}

\title[$HgSe_{0.7}S_{0.3}$ at High Pressure]{The pseudo-binary mercury
chalcogenide alloy HgSe$_{0.7}$S$_{0.3}$ at high pressure: a
mechanism for the zinc blende to cinnabar reconstructive phase
transition}

\author{D.P. Kozlenko\dag \footnote[3]{To whom correspondence should be addressed
(denk@nf.jinr.ru)}, K. Knorr\ddag, L. Ehm\ddag, S. Hull$\|$, B.N.
Savenko\dag, V.V. Shchennikov* and V.I. Voronin*}

\address{\dag\ Frank Laboratory of Neutron Physics, JINR, 141980
Dubna Moscow Reg., Russia}

\address{\ddag\ Institut f\"ur Geowissenschaften,
Mineralogie/Kristallographie, Universit\"at Kiel, Olshausenstr.
40, D-24098 Kiel, Germany}

\address{$\|$\ ISIS Facility, RAL, Chilton, Didcot, OX11 0QX, Oxon,
United Kingdom}

\address{*\ Institute for Metal Physics, Ural Branch of RAS,
620219 Ekaterinburg, Russia}

\begin{abstract}
The structure of the pseudo-binary mercury chalcogenide alloy
HgSe$_{0.7}$S$_{0.3}$ has been studied by  X-ray and neutron
powder diffraction at pressures up to 8.5 GPa. A phase transition
from the cubic zinc blende structure to the hexagonal cinnabar
structure was observed at $P \sim$ 1 GPa. A phenomenological model
of this reconstructive phase transition based on a displacement
mechanism is proposed. Analysis of the geometrical relationship
between the zinc blende and the cinnabar phases has shown that the
possible order parameter for the zinc blende - cinnabar structural
transformation is the spontaneous strain $e_{4}$. This assignment
agrees with the previously observed high pressure behaviour of the
elastic constants of some mercury chalcogenides.

\end{abstract}



\maketitle

\section{Introduction}

II-VI pseudo-binary solid solutions of the AB$_{1-x}$C$_{x}$ type
(A = Hg, Cd, Zn; B, C = Se, S, Te) have important applications in
heterojunctions and opto-electronic devices \cite{Letardi87}.
Knowledge of the structural, transport and thermodynamic
properties of these alloys is important for both the manufacturing
technology and the search for additional potential applications.

It has recently  been shown that pseudo-binary mercury
chalcogenide alloys HgSe$_{1-x}$S$_{x}$ exhibit an electronic
semimetal to semiconductor phase transition under high pressure
and that the transition pressure strongly depends on the sulfur
content $x$ \cite{Sh95,Sh97,Sh98}. In a subsequent structural
study it has been established that HgSe$_{1-x}$S$_{x}$ alloys with
0.3 $< x <$ 0.6 crystallise in the cubic zinc blende structure at
ambient conditions and the previously observed electronic phase
transition corresponds to a structural transformation to the
hexagonal cinnabar structure at $P \sim$ 1 GPa \cite{Voronin01}.

Cinnabar is the ambient pressure phase of HgS (cinnabar)
\cite{Auri50} and it has been found under high pressure in the
chalcogenides  HgSe \cite{Mariano63,McMahon96a}, HgTe
\cite{Werner83,Wright93,San95}, CdTe \cite{McMahon93}, ZnTe
\cite{Nelmes94} and ZnSe \cite{Porres01}, respectively. The role
of the cinnabar phase as an intermediate structure between the
fourfold coordinated zinc blende and sixfold coordinated rock salt
phases in II - VI compounds has extensively been studied both
experimentally \cite{Wright93, San95, Nelmes94, Porres01} and
theoretically \cite{Lu89, Cote97, Lee96, Qteish01} in the
recent years. In the cinnabar structure (space group $P3_{1}21$)
Hg (Cd, Zn) atoms occupy sites 3(a) ($u$, 0, 1/3) and chalcogen
atoms occupy sites 3(b) ($v$, 0, 5/6). Depending on the values $u$
and $v$, the structure of the cinnabar phase may exhibit different
atomic arrangements with coordination numbers 2+4 (HgS), 4+2
(HgTe) and 6. The latter corresponds to the rock salt  structure in
its hexagonal setting, having a lattice parameter ratio $c/a$ =
$\sqrt{6}$ and $u=v$=2/3) \cite{Wright93,Porres01}.

One should note that most of the previous theoretical studies of
the phase transitions in mercury chalcogenides have been based on
total energy ab-initio calculations. An alternative
approach is the analysis of the symmetry changes during the phase
transition and a consideration of the phenomenological model of
the phase transition in the framework of Landau theory using an order
parameter formalism \cite{Landau,Brus}.

Under high pressure a decrease of the elastic constant $C_{44}$
and the elastic constant combination $\frac{1}{2}$
($C_{11}$-$C_{12}$) has been observed in HgSe and HgTe in the
vicinity of the zinc blende - cinnabar phase transition
\cite{Ford82, Miller81}. Consequently, a spontaneous strain of
appropriate symmetry would be a primary order parameter for this
phase transition.

 The structure of the pseudo-binary mercury
chalcogenide alloys HgSe$_{1-x}$S$_{x}$ has recently been studied
\cite{Voronin01} over a limited pressure range up to 3 GPa and no
information on the pressure evolution of the cinnabar phase or
other possible phase transitions has been obtained. In this work
we studied the structural behaviour of the pseudo-binary alloy
HgSe$_{0.7}$S$_{0.3}$ in the extended pressure range up to 8.5
GPa. The resultant structural data were used for the discussion of
the zinc blende - cinnabar phase transition in the framework of
the Landau theory for phase transitions \cite{Landau, Brus} and
the search for a possible order parameter.

\section{Experimental}

Polycrystalline samples of HgSe$_{0.7}$S$_{0.3}$ were prepared by
melting of high purity components HgSe and HgS (99.999 \%). The
chemical composition of the sample was determined by means of
X-ray fluorescence analysis using a "Superprobe-JCXA-733"
spectrometer. The sulfur content was determined to be
0.302(1).

X-ray powder diffraction measurements were performed at pressures
up to 5 GPa with a Mar-2000 image plate diffractometer
($MoK_{\alpha}$ radiation, Si-(111) monochromator, $\lambda$ =
0.7107 \AA) using a Merrill-Bassett type diamond-anvil cell
 \cite{Merrill74}. The pressure transmitting medium was a 4:1
methanol - ethanol mixture and the pressure was measured using the
ruby fluorescence technique \cite{Pier75}. The two-dimensional
powder diffraction patterns were integrated by applying the FIT2D
program \cite{Hammer96} to give one dimensional conventional
powder diffraction profiles.

Neutron powder diffraction experiments at pressures up to 8.5 GPa
were performed at the POLARIS diffractometer \cite{Hull92} (ISIS,
RAL, UK), using the Paris - Edinburgh high pressure cell
\cite{Loveday96}. The sample volume was $V \sim$ 100 $mm^{3}$. The
scattering angle range was 2$\theta$ = 84-96$^{\circ}$ and the
diffractometer resolution in this geometry is approximately
$\Delta$$d/d \approx$ 0.007. Pressure were determined using the
compressibility data obtained from the X-ray diffraction
measurements up to 5 GPa and their extrapolation to higher
pressures. The experimental data were corrected for the effects of
attenuation of the incident and scattered beams by the pressure
cell components \cite{Wilson95}.

Due to the high absorption of X-rays and neutrons by the sample
typical exposure times were 16-18 h for both experiments.

\section{Results}

\begin{figure}
\begin{center}
\includegraphics{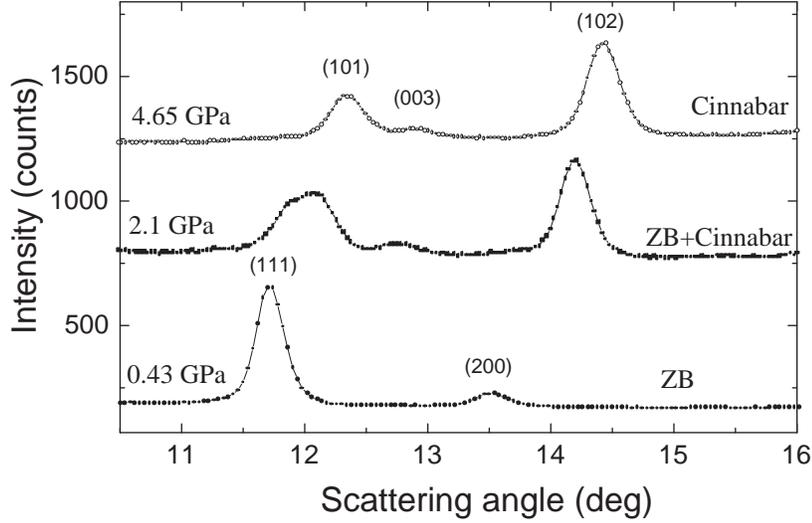}
\caption{\label{X-spec} Sections of the integrated X-ray powder
diffraction patterns of HgSe$_{0.7}$S$_{0.3}$ at pressures of
0.43, 2.1 and 4.65 GPa illustrating the evolution of the cinnabar
phase with increasing pressure.}
\end{center}
\end{figure}

Integrated X-ray powder diffraction patterns of
HgSe$_{0.7}$S$_{0.3}$ obtained at three different pressures are
shown in fig. \ref{X-spec}.
The pattern obtained at $P$ = 0 corresponds to the cubic zinc
blende structure. Peaks arising from the  hexagonal cinnabar
phase were first observed at $P$ = 0.97 GPa. The contribution from
the cinnabar phase increased with pressure and the single cinnabar
phase was observed at pressures above 2.1 GPa.

Refinement of the diffraction data in profile matching mode using
the Fullprof program \cite{Fullp93} provided the lattice
parameters of the zinc blende and cinnabar phases as functions of
pressure (table \ref{T1} ).

\begin{table}
\caption{\label{T1}Lattice parameters of the cubic zinc blende
($a_{cub}$) and the hexagonal cinnabar ($a$, $c$) phases of
HgSe$_{0.7}$S$_{0.3}$ as functions of pressure.}

\begin{indented}
\lineup
\item[]\begin{tabular}{@{}*{4}{l}}
\br
Pressure, GPa &
$a_{cub}$, \AA\ & $a$, \AA\ & $c$, \AA\cr
 \mr
 0 & 6.024(4) & - & - \cr
0.43(2) & 6.020(4) & - & -\cr
0.76(2) & 6.008(4) & - & -\cr
0.97(2) & 5.997(4) & - & -\cr
 1.68(3) & 5.975(4) & 4.167(4) & 9.605(4)\cr
2.10(3) & 5.945(4) & 4.146(4) & 9.585(4)\cr
 3.15(3) & - & 4.108(4) & 9.536(4)\cr
 3.78(3) & - & 4.087(4) & 9.524(4)\cr
4.65(4) & - & 4.066(4) & 9.497(4)\cr
\br
\end{tabular}
\end{indented}
\end{table}

\begin{figure}
\begin{center}
\includegraphics{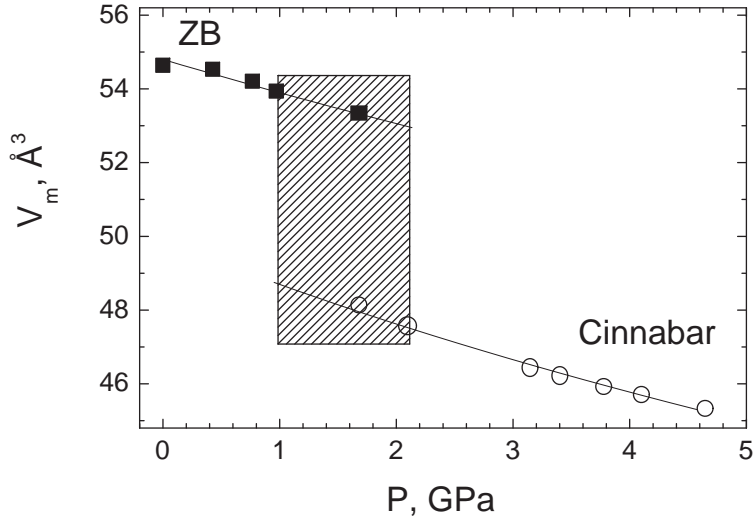}
\caption{\label{VM} The molar volume of HgSe$_{0.7}$S$_{0.3}$ as a
function of pressure. Error bars are within the size of the
symbols. The shaded area corresponds to the region of the
coexistence of the zinc blende  and the cinnabar phases. The solid
lines represent fits of a 2nd order  Birch-Murnaghan
equation-of-state   to the experimental data.}
\end{center}
\end{figure}

The pressure dependence of the molar volume ($V_{m}$) of
HgSe$_{0.7}$S$_{0.3}$ is shown in fig. \ref{VM}. The zinc
blende-cinnabar phase transition results in a volume discontinuity
of $\Delta$$V_{m}/V_{m0}$ $\approx$ 9 \%.

 For the description of the
compressibility of crystals, the 3rd order Birch-Murnaghan
equation-of-state \cite{Birch78} is commonly used
\begin{equation}
P=\frac{3}{2}B_0\left (x^{-7/3}-x^{-5/3}\right )\left
[1+\frac{3}{4}\left (B_1-4\right )\left (x^{-2/3}-1\right)\right],
\label{Eq1}
\end{equation}
where $x$ = $V/V_{0}$ = $V_{m}/V_{m0}$ is the relative volume
change, $V_{0}$ ($V_{m0}$) is the unit cell volume (molar volume)
at $P$ = 0 and $B_{0}$ and $B_{1}$ are the bulk modulus and its
pressure derivative. In Eq. \ref{Eq1} the pressure $P$ acts as the
response variable. Since volume as well as pressure are subject to
experimental uncertainty these errors were transformed into
effective variances of the pressure $\sigma_P^{\rm eff}$
\cite{Angel00}:
\begin{equation} \label{eq:weights}
 \sigma_P^{\rm eff} = \sqrt{\sigma_P^2 + (\sigma_V B_0/V)^2}.
\end{equation}
These effective variances were used as weights $w=1/(\sigma_P^{\rm
eff})^2$ in the least squares refinement of the equation-of-state.
As the bulk modulus $B_0$ appears in the right hand side of Eq.
\ref{eq:weights} weights were re-evaluated at each iteration step
of the least squares procedure. Since the compressibility data for
the zinc blende and the cinnabar phases were obtained in a rather
limited pressure range, it is difficult to obtain $B_{0}$ and
$B_{1}$ simultaneously. Hence, values of $B_{0}$ and $V_{m0}$ at
$P$ = 0 for the zinc blende and the cinnabar phases (table
\ref{T2}) were obtained from the fit of the experimental data by
the 2nd order Birch-Murnaghan equation-of-state which corresponds
to Eq. \ref{Eq1} with  $B_{1}$ = 4.

The calculated values of $B_{0}$  are higher than the corresponding
values obtained for the binary compounds HgSe and HgS (table
\ref{T2}). One would expect the increase of the bulk modulus for
the zinc blende phase of HgSe$_{1-x}$S$_{x}$ alloys in comparison
with HgSe because partial substitution of Se atoms by S atoms
decreases the average ionic radius of the $X$ = Se/S anion. It is more
difficult to compare values of $B_{0}$ for the cinnabar phases of
HgSe$_{1-x}$S$_{x}$ and HgS, since they exist in different
pressure ranges. However, the large difference between $B_{0}$ for
HgSe$_{1-x}$S$_{x}$ and HgS may be connected with the large value
of $B_1$ = 11.1 used for a description of the compressibility data
in \cite{Werner83}. It is well known that $B_{0}$ and $B_{1}$ are
correlated parameters. Therefore it is not surprising to find
smaller values for $B_{0}$ if $B_{1}$ is larger than 4 and vice
versa. A similar situation was found for another mercury
chalcogenide - HgTe \cite{San95}. For the cinnabar phase of this
compound the values $B_{0}$ = 41(10) GPa and $B_{1}$ = 3.3(2) were
obtained whilst in an earlier work a much smaller value $B_{0}$ = 16 GPa
was obtained with $B_{1}$ = 7.3.

In addition, the pressure dependence of the lattice
parameters for the cinnabar phase has been analysed. The volume in
Eq. \ref{Eq1} was substituted  by the cube of the lattice
parameters and fitted using the 2nd order Birch-Murnaghan
equation-of-state. The zero pressure compressibility $k_{a,c}$ is
then given by $1/(3B_0)$. The compressibility along the $a$-axis
is $k_a=0.0117$ GPa$^{-1}$ and  $k_c=0.0045$ GPa$^{-1}$ along the
$c$-axis. Hence, the compressibility in the cinnabar phase is
anisotropic with $k_a \approx 2.6 k_c$.

\begin{table}
\caption{\label{T2}The bulk modulus and molar volume at ambient
pressure for the zinc blende and cinnabar phases of
HgSe$_{0.7}$S$_{0.3}$, HgSe and HgS.}

\begin{indented}
\lineup \item[]\begin{tabular}{@{}*{5}{l}} \br
 & Zinc Blende & & Cinnabar &\cr
& HgSe$_{0.7}$S$_{0.3}$ & HgSe (Ref. 21) & HgSe$_{0.7}$S$_{0.3}$ &
HgS (Ref. 9)\cr
 \mr
$B_{0}$, GPa & 58(7) & 51.6(2) & 39(2) & 19.4(5)\cr $B_{1}$, GPa &
4 & 2.60(6) & 4 & 11.1\cr $V_{m0}$, \AA$^{3}$ & 54.8(1) & 56.3
(Ref. 17) & 49.9(2) & 47.2\cr \br
\end{tabular}
\end{indented}
\end{table}

\begin{figure}
\begin{center}
\includegraphics{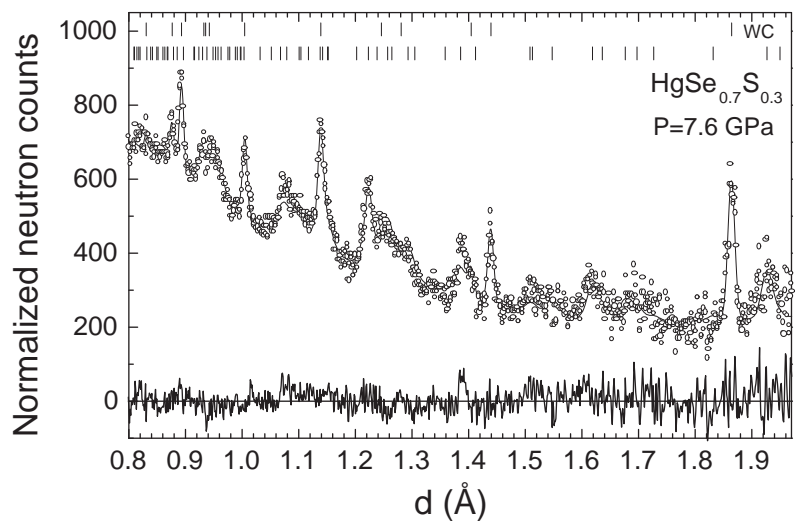}
\caption{\label{N-spec}The neutron diffraction pattern of the
cinnabar phase of HgSe$_{0.7}$S$_{0.3}$ measured at the POLARIS
diffractometer at $P$ = 7.6 GPa and processed by the Rietveld
method. Experimental points, calculated profile and difference
curve (below) are shown. The contribution from the tungsten
carbide (WC) anvils of the high pressure cell was also included in
the calculations.}
\end{center}
\end{figure}

It was not possible to obtain atomic positional parameters for the
cinnabar phase of HgSe$_{0.7}$S$_{0.3}$ from the X-ray diffraction
data due to the peak overlap between the reflections of the sample
and the gasket, the strong absorption by the sample and the
restricted $Q$~-~range in the X-ray experiment. However, this
information could be obtained from the neutron diffraction
experiments. The neutron diffraction pattern of the cinnabar phase
of HgSe$_{0.7}$S$_{0.3}$ measured at the POLARIS diffractometer at
$P$ = 7.6 GPa is shown in fig. \ref{N-spec}.

The diffraction data were refined by the Rietveld method using the
program MRIA \cite{Zlok92}. In the refinement procedure the
structural model \cite{Wright93} in space group $P3_{1}21$ was
used, with Hg atoms occupying 3($a$) sites ($u$, 0, 1/3) and $X$
atoms ($X$ = Se, S) occupying 3($b$) sites ($v$, 0, 5/6), as found
for the binary compounds $HgX$.
 The contribution to the diffraction
pattern from the tungsten carbide anvils of the high pressure cell
was treated as a second phase. The lattice parameters and atomic
positional parameters for the cinnabar phase of
HgSe$_{0.7}$S$_{0.3}$,  the calculated nearest-neighbour
interatomic distances and the obtained $R$-factors values at
different pressures are given in table \ref{Str-par}, along with
the corresponding values for the binary compounds HgSe and HgS. No
further phase transitions were observed in the investigated
pressure range up to 8.5 GPa.

\begin{table}
\caption{\label{Str-par}Refined structural parameters of the
 cinnabar phase of HgSe$_{0.7}$S$_{0.3}$ at different pressures.
 Lattice parameters ($a$, $c$); positional parameters of Hg and $X$ = Se/S
atoms ($u$, $v$) and the nearest-neighbour distances ($Hg1-X$,
$Hg2-X$, $Hg3-X$) are presented. The corresponding values for the
binary compounds HgSe and HgS are given for comparison.}

\begin{indented}
\lineup \item[]\begin{tabular}{@{}*{6}{l}} \br & &
HgSe$_{0.7}$S$_{0.3}$ & & HgSe \cite{McMahon96a} & HgS
\cite{Wright93}\cr $P$, GPa & 2.7(3) & 7.6(5) & 8.5(6) & 2.25 &
0\cr $a$, \AA\ & 4.127(5) & 3.982(5) & 3.957(5) & 4.174(1) &
4.14\cr $c$, \AA\ & 9.551(9) & 9.389(8) & 9.333(8) & 9.626(1) &
9.49\cr $u$ & 0.672(7) & 0.676(7) & 0.679(7) & 0.666(1) &
0.720(3)\cr $v$ & 0.529(5) & 0.558(5) & 0.572(5) & 0.540(1) &
0.480(10)\cr $Hg1-X$, \AA\ & 2.49(2) & 2.49(2) & 2.50(2) &
2.541(4) & 2.36(5)\cr $Hg2-X$, \AA\ & 2.93(2) & 2.84(2) & 2.82(2)
& 2.941(4) & 3.10(5)\cr $Hg3-X$, \AA\ & 3.28(2) & 3.08(2) &
3.01(2) & 3.299(5) & 3.30(5)\cr $R_{p}$, \% & 10.7 & 10.6 & 9.0\cr
$R_{wp}$, \% & 9.0 & 10.7 & 8.2\cr
 \br
\end{tabular}
\end{indented}
\end{table}

\section{Discussion}

The cinnabar structure is intermediate  between the fourfold
coordinated zinc blende and the sixfold coordinated NaCl-type
structure \cite{San95}. The zinc blende - cinnabar phase
transition is of the reconstructive type since these two
structures are not related by a direct group - subgroup
relationship \cite{Toledano96}.

\begin{figure}
\begin{center}
\includegraphics{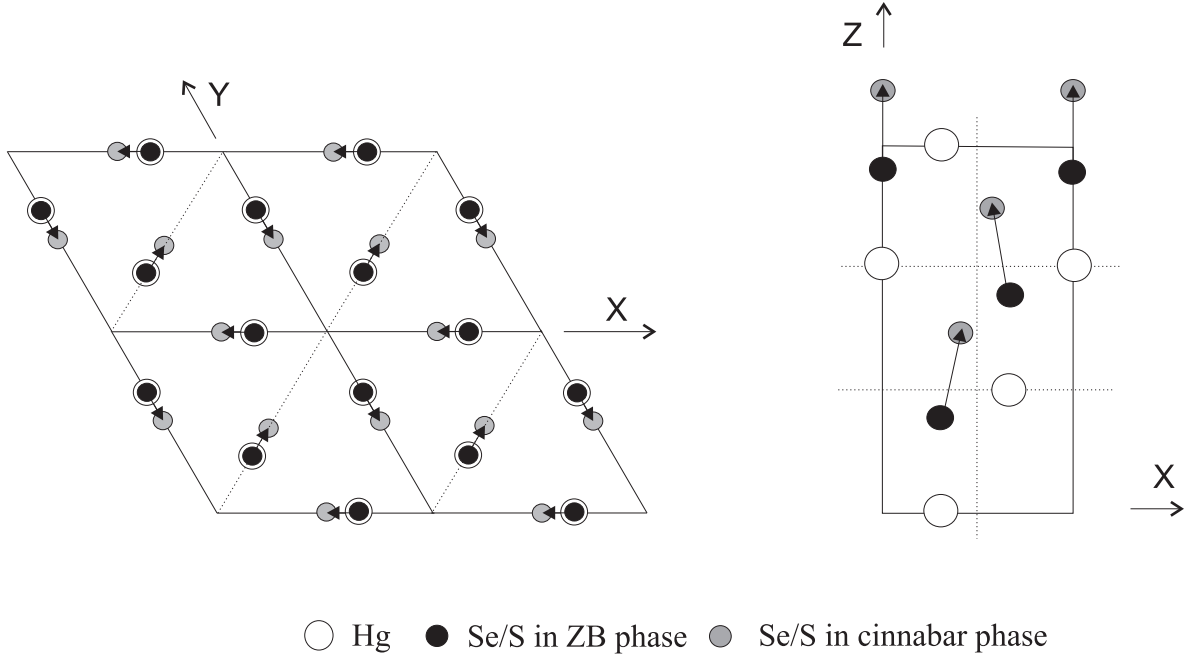}
\caption{\label{ZINC BLENDE-hex}Relationship between the  zinc
blende structure in the hexagonal setting and the cinnabar
structure. Projections of  the two structures on the $XY$ plane
(left) and on the $XZ$ plane (right) are shown. Displacements of
the atoms due to the transition are indicated by arrows.}
\end{center}
\end{figure}

For the analysis of the atomic displacements at the phase transition,
the zinc blende phase of HgSe$_{0.7}$S$_{0.3}$ may be described in
a hexagonal setting (fig. \ref{ZINC BLENDE-hex}) using space group
$P3_{1}$ with Hg and $X$ = Se/S atoms occupying sites 3($a$) ($x$,
$y$, $z$) (table \ref{T4}). The characteristic feature of the zinc
blende - cinnabar phase transition is a displacement of the $X$
atoms from their initial positions 3($a$) in space group $P3_{1}$
along the $z$ - direction by 1/4 $c$ and along the $x$ and $y$ -
directions by $\varepsilon$ $\sim$ 0.1 $a$ to the positions 3($b$)
of the space group $P3_{1}21$. The Hg atoms remain at nearly the
same positions 3($a$) in both phases (tables \ref{Str-par},
\ref{T4}, fig. \ref{ZINC BLENDE-hex}).

Assuming a displacement mechanism for the transition, the
transformation path from the zinc blende to the cinnabar phase can be
decomposed into three successive steps:
\begin{enumerate}
\item a distortion of the unit cell with the related displacement
of the atoms which lowers the cubic $F\bar 43m$ zinc blende
symmetry to $R3m$ symmetry;
 \item a displacement of the atoms in the $z$-direction of the rhombohedral lattice
  which lowers further the symmetry to $R3$. This $R3$ symmetry with one
  molecule per unit cell can be reseted in the hexagonal basis as
  the $P3_1$ space group with $Z=3$;
  \item an increase of the symmetry from $P3_1$ to $P3_121$.
\end{enumerate}

This displacement mechanism is fully reversible and corresponds
to Brillouin zone-centre phonon modes which involve simple
deformations of the initial cubic unit cell. The assumed
intermediate states display the polar symmetries $R3m$ and $R3$
(or $R3_{1}$), i.e. they are ferroelectric.

Among the macroscopic quantities which correspond to the symmetry
breaking mechanisms at the first step of the proposed
transformation path ($F\bar43m\rightarrow R3m$) spontaneous strain
components $e_{yz}$, $e_{xz}$, $e_{xy}$ appear. The symmetry
breaking transition from $F\bar43m$ to $R3m$ is proper
ferroelastic and, according to Janovec et al. \cite{Janovec75}, the
Landau condition is not fulfilled for the associated irreducible
representation of the $F\bar43m\rightarrow R3m$ symmetry change,
i.e. a cubic term appears in the corresponding free energy
expansion
\begin{equation}
F = \frac{\beta_{0}(P-P_{c})}{2}Q^2 + \frac{\gamma}{3}Q^3 +
\frac{\delta}{4}Q^4.
\end{equation}
The stability condition is
\begin{equation}
 \frac{\partial F}{\partial Q} = Q\left [ \beta_{0}(P-P_{c}) + \gamma Q + \delta Q^2\right ] = 0
\end{equation}

The order parameter $Q$ should have the symmetry of the strain. If
we consider a primitive rhombohedral lattice corresponding to the
zinc blende and cinnabar structures, the geometry of the
displacements of the $X=$ Se/S atoms due to the zinc blende -
cinnabar phase transition allows us to assume that a spontaneous
strain $e_4= e_5 = e_6$ ($e_{yz} = e_{xz} = e_{xy}$) could be a
possible order parameter.

Hence the pressure evolution of the spontaneous strain may be
described as
\begin{equation}
e_4=\frac{-\gamma}{2\delta}(1+\sqrt{1-4\beta_0(p-p_c)\delta/\gamma^2}).
\label{Eq3}
\end{equation}

\begin{figure}
\begin{center}
\includegraphics{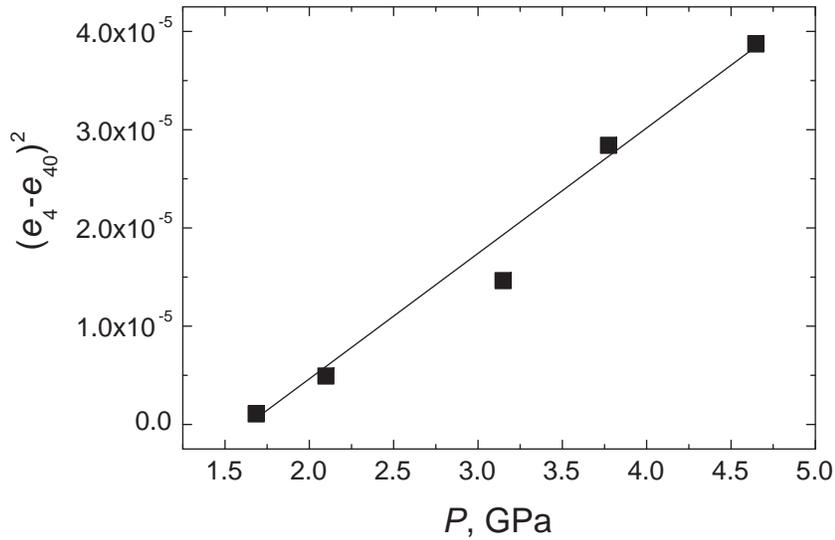}
\caption{\label{E52} Square of the relative spontaneous strain
($e_{4}-e_{40}$) as a function of pressure. The solid line is the
linear fit to the experimental data. Error bars are within the
symbols size.}
\end{center}
\end{figure}

Spontaneous strain $e_4$ stems from the distortion of the lattice
angle $\alpha$ of the primitive rhombohedral unit cell of the
cinnabar structure in comparison with its initial value $\alpha_0
= 60^o$ corresponding to the cubic fcc zinc blende structure. At
the first order zinc blende - cinnabar phase transition point,
$e_4$ undergoes a negative jump $e_{40}$ resulting in an
increasing of $\alpha$ from $60^o$ to $62.8^o$ and afterwards it
starts to increase with pressure increase reflecting the decrease
of $\alpha$ from $62.8^o$ to $62.1^o$. Fig. \ref{E52} shows the
square of the relative spontaneous strain ($e_4$-$e_{40}$) as a
function of pressure. It increases linearly with the pressure
increase, in agreement with Eq. \ref{Eq3}. To estimate $e_{40}$,
values of the ($a$, $c$) lattice parameters of the cinnabar
structure extrapolated to the estimated transition pressure $P_c$
= 0.97 GPa were used and the value $e_{40}$ = 0.025 was obtained.

\begin{table}
\caption{\label{T4}Atomic coordinates of the zinc blende structure
described in the hexagonal setting and the cinnabar structure
(values of positional parameters obtained at $P$ = 8.5 GPa are
given).}

\begin{indented}
\lineup \item[]\begin{tabular}{@{}*{4}{l}} \br
 $X$ = Se/S atoms & & Hg atoms & \cr
 \mr
Zinc Blende - 3($a$) sites & Cinnabar - 3($b$) sites & Zinc Blende
- 3($a$) sites & Cinnabar - 3($b$) sites\cr (2/3, 0, 7/12) &
(0.57, 0, 5/6) & (2/3, 0, 1/3) & ($\approx$2/3, 0, 1/3)\cr (0,
2/3, 11/12) & (0.57, 0, 1/6) & (0, 2/3, 2/3) & (0, $\approx$2/3,
2/3)\cr (1/3, 1/3, 1/4) & (0.43, 0.43, 1/2) & (1/3, 1/3, 0) &
($\approx$1/3, 1/3, 0)\cr
 \br
\end{tabular}
\end{indented}
\end{table}

Assuming  the spontaneous strain $e_{4}$  to be a primary order
parameter for the zinc blende - cinnabar phase transition, a
softening of the elastic constant $C_{44}$ in the vicinity of the
transition point should be expected \cite{Cowley76}. In the
high-pressure studies of the elastic constants of HgSe
\cite{Ford82} and HgTe \cite{Miller81}  a decrease of the $C_{44}$
has been observed on approaching  the zinc blende - cinnabar
transition pressure. This is in  agreement with our
considerations.

The cinnabar structure is closely related to the cubic NaCl-type
structure \cite{San95}. The NaCl-type structure in its hexagonal
setting may be described using the same space group as for the
cinnabar structure, $P3_{1}21$, with atomic positional parameters
$u$ = $v$ = 2/3. As shown in table \ref{Str-par}, with increasing
pressure the positional parameter of the Hg atoms varies slowly
and remains close to its value in the zinc blende structure or the
corresponding value for the NaCl-type structure, $u \sim$ 0.67.
The positional parameter of the $X$ = Se/S atoms (which shifts
from $v$ = 0.67 to $v \sim$ 0.53 at the phase transition,
increases up to 0.57 in the pressure range 2.7 - 8.5 GPa, rending
towards the value corresponding to the NaCl-type cubic structure.

\begin{figure}
\begin{center}
\includegraphics{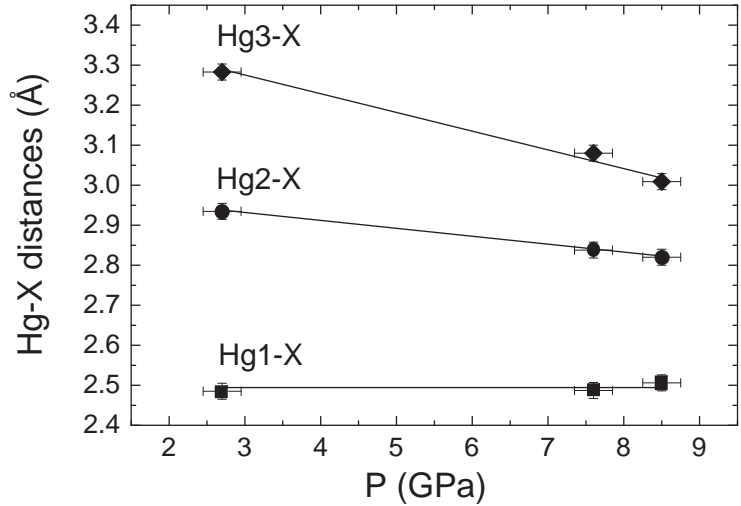}
\caption{\label{Hg-X}The nearest-neighbour distances between Hg and
$X$ = Se/S atoms in HgSe$_{0.7}$S$_{0.3}$ as functions of
pressure. The solid lines are linear interpolations of the
experimental data.}
\end{center}
\end{figure}

There are  three pairs of nearest-neighbour distances between $X$
= Se/S and Hg atoms ($Hg1-X$, $Hg2-X$ and $Hg3-X$) with rather
close values in the cinnabar structure \cite{Wright93}. In
HgSe$_{0.7}$S$_{0.3}$ the pressure increase from 2.7 to 8.5 GPa
causes the shortest $Hg1-X$ distance to remain almost constant
whilst the two other distances decrease approaching the value of
the first one (fig. \ref{Hg-X}, table \ref{Str-par}). This
corresponds to an increasing of the $X-Hg-X$ angle (from
168.1$^{\circ}$ to 173.8$^{\circ}$) and a decrease in the
 $Hg-X-Hg$ angle from 105.4$^{\circ}$ to
99.0$^{\circ}$. Thus, the cinnabar structure of
HgSe$_{0.7}$S$_{0.3}$ approaches the NaCl-type cubic structure
with increasing pressure, where the distances $Hg1-X$ = $Hg2-X$ =
$Hg3-X$ and the interatomic $X-Hg-X$ and $Hg-X-Hg$ angles have
values of 180$^{\circ}$ and 90$^{\circ}$, respectively.

A description of the coordination of the cinnabar phase of
HgSe$_{0.7}$S$_{0.3}$ may be performed using the following ratio:
    $r$ = ($l_{Hg2-X}$ - $l_{Hg1-X}$)/($l_{Hg3-X}$ - $l_{Hg2-X}$).
Values of $r <$ 1 correspond to a 4+2 coordination, $r >$ 1 to a
2+4 coordination and $r$ = 1 to the ideal sixfold coordination. The
calculated value $r$ for HgSe$_{0.7}$S$_{0.3}$ varies from 1.26 to
1.68 as the pressure increases from 2.7 to 8.5 GPa. The
increase of the distortion index $r$ shows that the difference
between the $Hg3-X$ and $Hg2-X$ distances decreases faster than
the difference between distances $Hg2-X$ and $Hg1-X$ on pressure
increase. A comparison between HgSe$_{0.7}$S$_{0.3}$ and the
parent binary compounds HgSe and HgS shows that the cinnabar
structure of HgSe$_{0.7}$S$_{0.3}$ is similar to that of HgSe ($r$
= 1.1, calculated from data reported in \cite{McMahon96a}) and
less distorted with respect to the NaCl - type cubic structure
($r$ = 1) than the cinnabar structure of HgS ($r$ = 3.7)
\cite{Wright93}. This means that the phase transition between the
cinnabar and NaCl-type phases in HgSe$_{0.7}$S$_{0.3}$ might be
expected to occur at roughly the same pressure as for HgSe, $P
\sim$ 16 GPa. In \cite{Sh93} a decrease of the electrical
resistivity by several orders of magnitude was observed in
HgSe$_{0.7}$S$_{0.3}$ at $P \sim$ 15 GPa. Such a behaviour
corresponds to an electronic semiconductor-metal phase transition
which accompanies the structural phase transition from the
cinnabar to the NaCl-type structure in binary mercury
chalcogenides HgSe, HgTe \cite{Huang85,San95}.

\section{Conclusions}

For the first time the pressure-induced zinc blende - cinnabar phase
transition has been analysed in terms of atomic displacements
during the reconstructive phase transition. The proposed
phenomenological model of the phase transition includes three
successive steps with a symmetry change $F4\bar3m\rightarrow
R3m\rightarrow P3_1\rightarrow P3_121$.

The analysis of the geometry of the atomic displacements shows
that a possible order parameter for the zinc blende - cinnabar
structural transformation is a spontaneous strain $e_{4}$. This
conclusion is
 supported by a softening of the elastic constant $C_{44}$ at this
phase transition as observed in HgSe \cite{Ford82} and HgTe
\cite{Miller81}.

The proposed model of the zinc blende - cinnabar phase transition
has a common character and may be generalised for the case of
other mercury chalcogenides exhibiting this phase transition -
HgSe, HgTe, CdTe, ZnTe and their pseudo-binary solutions.

The coordination and geometrical features of the cinnabar phase of
HgSe$_{0.7}$S$_{0.3}$ are similar to those of HgSe. With
increasing pressure the interatomic angles and distances of the
cinnabar structure approach the values corresponding to the
NaCl-type cubic structure. We therefore expect  a phase transition
from the hexagonal cinnabar structure to the cubic NaCl-type
structure at higher pressure.

\ack {The authors are grateful to Prof. P.W. Toledano for the
helpful discussion of the phase transition mechanism. The work was
supported by the Russian Foundation for Basic Research, grants
00-02-17199 and 01-02-17203.}

\pagebreak
\end{document}